# High-capacity computing with self-rectification nonlinear optical neural processor

Ruicheng Ma[1,2,3,4#], Siyu Dong[1,2,3,4#], Yuzhi Shi[1,2,3,4#*], Yuchen Zhu[1,2,3,4], Hong Luo[1,2,3,4], Qiang Fu[5], Hadi Amata[5], Wolfgang Heidrich[5], Xiong Dun[1,2,3,4], Hongfei Jiao[1,2,3,4], Hui Zhang[1,2,3,4], Qinghua Song[6], Zeyong Wei[1,2,3,4*], Zhanshan Wang[1,2,3,4], Ali Momeni[7], Romain Fleury[7*] and Xinbin Cheng[1,2,3,4*]

[1] Institute of Precision Optical Engineering, School of Physics Science and Engineering, Tongji University, Shanghai 200092, China.

[2] MOE Key Laboratory of Advanced Micro-Structured Materials, Shanghai 200092, China.

[3] Shanghai Frontiers Science Center of Digital Optics, Shanghai 200092, China.

[4] Shanghai Professional Technical Service Platform for Full-Spectrum and High-Performance Optical Thin Film Devices and Applications, Shanghai 200092, China.

[5] King Abdullah University of Science and Technology (KAUST), Thuwal, Saudi Arabia.

[6] Institute of Materials Research (IMR), Tsinghua Shenzhen International Graduate School, Tsinghua University, Shenzhen, China.

[7] Laboratory of Wave Engineering, Department of Electrical Engineering, EPFL, Lausanne CH-1015, Switzerland.

#These authors contributed equally.

*Corresponding authors. Emails: yzshi@tongji.edu.cn (Y.S.); weizeyong@tongji.edu.cn (Z.W.); romain.fleury@epfl.ch (R.F.); chengxb@tongji.edu.cn (X.C.)

## Abstract

Artificial intelligence (AI) and neural networks have driven groundbreaking innovations across numerous disciplines. Optical computing offers the promise of unprecedented speed and energy efficiency in the post-Moore era; however, achieving efficient, practical nonlinear activation using all-optical approaches remains a challenge. Here, we present an optical nonlinear neural processing unit (ONNPU) that implements all-optical nonlinear activation through a self-rectification mechanism. The ONNPU architecture perfectly imitates the structure of digital neural networks, enabling seamless integration with the established deep learning ecosystem. We benchmark ONNPU across nine diverse tasks spanning decision, regression and generation, including accuracies of 98.07% on MNIST and 93.54% on Fashion-MNIST. When integrated into a 201-million-parameter Vision Transformer, ONNPU achieves 82.4% top-1 accuracy on full ImageNet classification (1,000 categories); when integrated into a 117-million-parameter decoder-only Transformer, ONNPU enables short-form story generation that outperforms GPT-2. By addressing more complex and diverse deep learning tasks, ONNPU paves the way toward practical optical machine intelligence, unleashing significant potential for high-performance optical computing.

**Keywords:** high-capacity optical computing, expressive nonlinearity, optical large language model, large-scale optical computing, all-optical nonlinear activation

## Introduction

Optical computing holds transformative potential for artificial intelligence [1–5], offering orders-of-magnitude improvements in computing speed and energy efficiency by exploiting the intrinsic parallelism and near-zero-dissipation dynamics of light propagation [6–12]. Despite substantial progress in implementing large-scale linear operations through diffractive layers [13–21], Mach-Zehnder interferometer meshes [22–24], and micro-rings [25–28], the lack of practical high-capacity all-optical nonlinear activation remains a formidable bottleneck to realizing deep optical networks with performance competitive with their electronic counterparts [29–32]. Consequently, most research has gravitated toward hybrid optoelectronic schemes, in which optical signals are detected, passed through digital circuits to apply activation functions, and re-encoded into light [9,10,33–40]. While such designs preserve optical parallelism and electronic programmability, the repeated conversion between domains imposes substantial latency and power overhead that ultimately constrains throughput [41].

Achieving the full performance gains of optical neural networks necessitates the transition to all-optical activation functions. To this end, research efforts have bifurcated into two primary paradigms. The first harnesses intrinsic nonlinear optical effects in materials—encompassing phenomena such as induced transparency [42,43], ferroelectric films [44], spatiotemporal Kerr effect [45,46], second-harmonic generation [47,48], saturable absorption [49], zero-mode lasing [50], and third-harmonic generation [51]—to elicit all-optical nonlinear responses. However, the practical utility of these physical processes is often constrained by stringent operating conditions or the requirement for high optical power, thereby posing significant hurdles for scalable, energy-efficient deployment. In addition, because the nonlinear response is often entwined with device physics and operating regimes, it can be difficult to expose a clean, reusable activation module that maps directly onto the layerwise "linear transform–activation" abstraction of electronic networks. Many of these effects also manifest as low-order polynomial intensity relations (for example, quadratic scaling familiar from second-order nonlinear optics), which can further tighten the trade-off between nonlinearity strength and peak power or wall-plug efficiency. Recently, complete photonic integrated neurons leveraging Kerr nonlinearity in micro-ring resonators have demonstrated impressive performance in spatiotemporal processing with ultralow latency [52], yet the integration scale remains limited by on-chip footprint constraints [53]. The second paradigm exploits structural nonlinearity [54–57], which encodes input data repetitively across multiple scattering events [54] or diffractive layers [55] to synthesize polynomial expansions at low optical power. Such architectures can be effectively trained for specific inference tasks and achieve enhanced accuracy. However, such data repetition leads to the loss of universal linear transformation capability and typically requires offline pre-computation, adding overhead beyond the layerwise forward pass. This route is therefore less naturally aligned with standard layer-by-layer training and resource accounting in mainstream deep learning, where each stage is ideally a universal linear map followed by a local nonlinearity. Along this line, an elegant variant enables massively parallel universal approximation via linear diffractive processing, yet still lacks end-to-end neural-network validation for strong ecosystem adaptability [58]. Thus, structural nonlinearity blocks cannot serve as optical analogs to the fully connected or convolutional layers commonly employed in digital neural networks [59]. It is useful to contrast these optical paradigms with what digital networks actually use in practice: depth is made effective by interleaving linear layers with element-wise, threshold-like nonlinearities in the ReLU family [29], while classical approximation theory underscores the role of non-polynomial, neuron-wise activations—including rectifying and threshold-like functions.

Here, we present an optical nonlinear neural processing unit (ONNPU) that addresses these limitations through a self-rectification mechanism. Crucially, the ONNPU architecture, which consists of diffractive linear layers interleaved with all-optical nonlinear activation, directly mirrors the canonical structure of digital neural networks (linear transform–activation). In intuitive terms, self-rectification is the optical counterpart of ReLU-like activations: at each spatial pixel, the field is strongly suppressed below an intensity threshold and transmitted once the threshold is exceeded, so the processor alternates "global linear mixing" with "local on/off gating"—the same pattern as linear layers followed by ReLU-type nonlinearities in electronic deep networks [29]. This enables seamless integration with established network topologies, training algorithms, and the broader deep learning ecosystem [60]. This architectural compatibility distinguishes ONNPU from approaches that require specialized network designs or unconventional training paradigms. Furthermore, the massively parallel nature of self-rectification allows computing speed to scale with the number of neurons ( $\propto N^2$) while the processing latency remains constant, providing a pathway to large-scale optical computing without proportional overhead in time or energy per operation. We experimentally demonstrate ONNPU across nine diverse deep learning benchmarks. On image classification, ONNPU achieves accuracies of 98.07% (MNIST), 93.54% (Fashion-MNIST) without electronic post-processing, and 91.24% (CIFAR-10), 81.71% (Imagenette, a 10-category ImageNet subset) with a linear electronic readout. By assembling multiple ONNPU modules into a vision transformer architecture [61], we further extend to 1000-category ImageNet classification (82.4%). Beyond vision, we integrate ONNPU modules into a 117M-parameter decoder-only, causal Transformer for short-form story generation, which outperforms GPT-2 (124M) under a double-blind swap-correction LLM-judge protocol. The system delivers an end-to-end computing speed of 590,000 tasks per second and energy efficiency of 11,000 tasks/J, with neuron counts scaling from $3.23\times10^6$ in a single module to $2.01\times10^8$ in the multi-module configuration. By establishing an optical nonlinear architecture fully compatible with mainstream neural network frameworks, ONNPU paves the way toward practical optical machine intelligence.

## Results

### The architecture of ONNPU

The ONNPU consists of diffractive linear layers interleaved with an all-optical nonlinear activation module (Fig. 1a). At the system level, ONNPU combines high architectural adaptability with competitive figure-of-merit performance across nonlinear optical computing paradigms (Fig. 1b–c). Input data encoded at the fundamental frequency $\omega$ propagate through cascaded diffractive phase layers that implement massively parallel linear transforms. The nonlinear activation is configured through a self-rectification module. First, a bulk nonlinear crystal converts the input laser into the double-frequency light ($2\omega$) via second-harmonic generation upon exceeding an intensity threshold. Then, a wavelength-selective filter transmits only the $2\omega$ component, effectively filtering out sub-threshold signals (Fig. 1d). This frequency-conversion mechanism yields a threshold-like transfer function that is implemented entirely in the optical domain, as verified by experimental measurements that align closely with theoretical predictions (Fig. 1e).

As the self-rectification operates simultaneously across all spatial neurons without electronic conversion, scaling the ONNPU incurs no additional latency penalty per neuron. This massive parallelism enables the computing speed to scale quadratically with neuron count while latency remains constant at sub-nanosecond levels, as shown in Fig. 1f. To ensure fair comparison with state-of-the-art systems, we evaluate ONNPU using a rigorous calculation method that treats cascaded linear layers as equivalent single-layer transformations [11] and ensures adequate intervals between inputs (Supplementary Information, Fig. S5). Based on these metrics, ONNPU experimentally achieved 590,000 tasks/s and 11,000 tasks/J in terms of systematic computing speed and energy efficiency, respectively (Supplementary Information, Note S1 and Table S1). These metrics are comparable to those of NVIDIA A100 [62,63], even with a non-top-tier SLM. Benchmarked against optical and electronic/electro-optic nonlinear systems (Fig. 1g–h and Table 1), ONNPU achieves competitive or superior performance across computing speed, energy efficiency, network scale, and accuracy.

The architectural consistency between ONNPU and conventional neural networks allows for the direct adoption of established training methods. We train diffractive layer parameters using standard backpropagation with an error-aware optical forward model (Supplementary Information, Note S6 and Fig. S1). Beyond in silico training, ONNPU is also compatible with emerging physical training paradigms such as physics-aware training [64–66] and fully forward mode onsite learning [36,67,68], which implement gradient computation directly on physical systems.

## Benchmarking ONNPU on image classification

We evaluate the classification performance of ONNPU across standard image recognition benchmarks. As illustrated in the experimental pipeline (Fig. 2a–b), input images are encoded onto the spatial light modulator (SLM), propagated through cascaded diffractive layers interleaved with self-rectification modules, and recorded by the camera for region-wise classification. To maintain stable multilayer registration in the free-space stack, the diffractive chips are mounted and aligned using a CNC (Computer Numerical Control) metal holder (Fig. 2c–e and Supplementary Information, Fig. S2), with the associated robustness-aware training method provided in Supplementary Information (Note S6 and Fig. S1).

ONNPU achieves 98.07% classification accuracy on the MNIST handwritten digit dataset [69] and 93.54% on the Fashion-MNIST dataset [70]. The training dynamics (Fig. 2f–g) demonstrate the critical role of self-rectification. Networks trained with self-rectification converge rapidly, reaching 98.07% accuracy on MNIST and 93.54% on Fashion-MNIST within the first few epochs. In contrast, networks without self-rectification plateau at substantially lower accuracy levels (~80%). The confusion matrix of ONNPU (Fig. 2h–i) also shows strong diagonal dominance with minimal inter-class confusion.

For higher-complexity datasets, ONNPU adopts a hybrid multi-pass configuration with a lightweight linear electronic readout with fewer than $10^3$ trainable parameters (Fig. 2j). Under this setting, ONNPU achieves 91.24% on CIFAR-10 [71] and 81.71% on Imagenette [72] (Fig. 2k–l), substantially outperforming representative optical baselines under comparable settings (Supplementary Information, Table S2).

**ONNPU on diverse input discrimination**

We evaluate ONNPU on two tasks requiring fundamentally different feature representations: geometric property recognition and music genre classification. The Kakeya membership classification task tests global geometric reasoning rather than local texture recognition. The Kakeya problem asks whether a unit-length line segment can rotate through 360° while remaining entirely within a given shape (Fig. 3a). We constructed four shape classes combining convexity and Kakeya membership (Fig. 3b). The training dynamics and confusion matrix (Fig. 3c–d) show rapid convergence to near-100% accuracy over 10 epochs, indicating learned recognition of global geometric properties associated with the rotational containment.

For music genre classification, audio signals of the FMA (Free Music Archive) dataset [73] from eight musical genres were converted to Mel-spectrograms and encoded as 2D optical intensity patterns (Fig. 3e–f). The training curve and confusion matrix (Fig. 3g–h) show rapid convergence, with accuracy exceeding 72.91% over 100 epochs. Notably, a t-SNE visualization reveals substantial overlap between genre features (Supplementary Information, Fig. S8), underscoring the intrinsic difficulty of this task and contextualizing the achieved performance.

These experiments demonstrate that ONNPU adapts from visual patterns to time-frequency representations and abstract geometric reasoning through the self-rectification nonlinearity, maintaining sub-nanosecond optical inference and high classification accuracy.

**ONNPU on regression**

We also evaluate ONNPU on regression tasks for continuous function fitting. We design three target functions motivated by real-world applications: industrial inspection detecting sudden state transitions (Fig. 4a), target acquisition in radar systems localizing events (Fig. 4b), and medical examination monitoring periodic physiological signals (Fig. 4c). These archetypes, specifically sharp thresholding, localized energy concentration, and high-frequency oscillation, represent complementary computational challenges.

With self-rectification, ONNPU achieves near-perfect fitting for all three target functions (Fig. 4d–f): $R^2$ = 0.999 and root mean square error RMSE = 0.001 for all functions. The predicted curves closely track the ground truth, including sharp transitions and fine-scale oscillations. In contrast, networks without self-rectification (Fig. 4g–i) exhibit substantially degraded performance: $R^2$ = 0.980, 0.951, 0.380, and RMSE = 0.065, 0.043, 0.089, respectively, failing to capture sharp discontinuities and high-frequency variations.

These results underscore the fundamental advantage of self-rectification: a purely linear optical network, regardless of its depth, collapses to a single linear mapping and struggles to fit complex nonlinear functions (Fig. 4j–l). The threshold-like activation enables the composition of nonlinear mappings, allowing ONNPU to effectively handle both classification and regression tasks.

### Full ImageNet classification with ONNPU ViT

To demonstrate ONNPU's capability on large-scale and complex tasks, we evaluate its performance on the full ImageNet dataset comprising 1,000 object categories, as shown in Fig. 5. We integrate ONNPU into a Vision Transformer (ViT) architecture by replacing the computationally intensive linear transformations and nonlinear activations in the embedding, encoder, and classifier stages with ONNPU modules, while retaining electronic computing for position encoding, attention matrix operations ($QK^\top$), and softmax normalization [7] (Fig. 5a and Supplementary Information, Fig. S6). This hybrid optical-electronic design leverages ONNPU's massive parallelism for the dominant matrix multiplications and nonlinear activations, achieving 81.8%–99.5% computational efficiency in the optical domain while maintaining full compatibility with the Transformer architecture (Supplementary Information, Note S8 and Table S3–S4). This benchmark represents a substantial leap in complexity compared to previous demonstrations, requiring discrimination across diverse visual categories spanning natural scenes, animals, objects, and abstract concepts.

The ONNPU Vision Transformer achieves 82.4% top-1 accuracy with 201M parameters, significantly outperforming vanilla optical neural networks (1.9% top-1 accuracy), as shown in Fig. 5b. It also exhibits competitive performance compared to state-of-the-art electronic models, including CoCa (91.0% top-1 accuracy with 2.1B parameters) [74] and ResNet-101 (86.4% top-1 accuracy with 829M parameters) [75]. The training dynamics in Fig. 5c show steady convergence over 75 epochs, achieving 82.4% top-1 accuracy and 96.5% top-5 accuracy. The top-5 metric, which considers a prediction correct if the true class appears among the top five predicted classes, demonstrates ONNPU's ability to capture semantically relevant features even when exact class identification is challenging. The progressive improvement in both metrics throughout training indicates effective learning of discriminative features across the full 1,000-class space (Fig. 5d–e). The overall performance of 82.4% top-1 and 96.5% top-5 accuracy on the full ImageNet dataset demonstrates that ONNPU can scale to large-scale and complex AI tasks while preserving the parallel and ultrafast computing capability inherent to optical processing.

### Language generation with ONNPU GPT

To extend the capability of ONNPU modules to generative language modeling, we embed them in a 117-million-parameter decoder-only, causal Transformer for short-form story generation (ONNPU GPT). As in the Vision Transformer co-design, ONNPU implements the dominant linear projections and self-rectification nonlinear activations across token embedding, fourteen decoder blocks, and the language-model head, while RMSNorm, attention-score computation ($QK^{\top}$), softmax, and sampling remain electronic (Fig. 6a and Supplementary Information, Fig. S7), yielding 81.8%–99.5% optical arithmetic share (Supplementary Information, Note S8).

ONNPU GPT produces continuations that compare favorably with electronic baselines (GPT-2, 124M; Llama-3, 8B) across structure, logic, fluency, alignment, and safety, with stable fidelity over story length and predominantly well-formed endings (Fig. 6b–j and Supplementary Information, Note S9). Under a double-blind swap-correction protocol with an independent LLM judge (sample number $n$ = 500), ONNPU GPT is preferred over GPT-2 in a majority of valid comparisons and remains competitive against Llama-3 (Fig. 6k–l). These results establish that self-rectification–based optical modules can underpin both Vision and decoder-only Transformers, extending spatial optical computing to language generation.

## Discussion

The ONNPU architecture preserves the canonical structure of digital neural networks, enabling seamless integration into the existing deep learning ecosystem. Alongside recent advances in optically assisted training of electronically implemented models [76], ONNPU places optics within the model's computational backbone, realizing the linear–nonlinear blocks of Vision and language Transformers, with 81.8%–99.5% of their arithmetic mapped to the optical domain. Our theoretical and experimental results demonstrate competitive performance across diverse AI tasks, with computing speed and energy efficiency exceeding those of state-of-the-art electronic and optical processors.

Looking forward, integrating self-rectification modules into chips leveraging metasurfaces or photonic integrated circuit technologies may enable more compact and scalable optical neural networks, thereby reducing system footprint and enhancing integration density [77,78]. Developing hybrid optical-electronic architectures that combine optical front-end processing with electronic memory and adaptive control could leverage the strengths of both computing paradigms, namely optical throughput for parallel operations and electronic precision for sequential logic and storage [79]. The architectural compatibility of ONNPU with mainstream neural network frameworks also suggests potential applications in generative models, where the massive parallelism and energy efficiency of optical computing could accelerate large-scale generation tasks [11,80,81] (Fig. 6 and Supplementary Information, Table S5). Self-rectification-based optical neural processing could serve as a foundation for practical optical machine intelligence systems in the post-Moore era.

## References

1. Wetzstein, G. et al. Inference in artificial intelligence with deep optics and photonics. *Nature* **588**, 39–47 (2020).

2. Shastri, B. J. et al. Photonics for artificial intelligence and neuromorphic computing. *Nat. Photonics* **15**, 102–114 (2021).

3. Zangeneh-Nejad, F., Sounas, D. L., Alù, A. & Fleury, R. Analogue computing with metamaterials. *Nat. Rev. Mater.* **6**, 207–225 (2021).

4. Wu, N. et al. Intelligent nanophotonics: when machine learning sheds light. *eLight* **5**, 5 (2025).

5. Bente, I. et al. The potential of multidimensional photonic computing. *Nat. Rev. Phys.* **7**, 439–450 (2025).

6. McMahon, P. L. The physics of optical computing. *Nat. Rev. Phys.* **5**, 717–734 (2023).

7. Anderson, M. G., Ma, S.-Y., Wang, T., Wright, L. G. & McMahon, P. L. Optical transformers. Preprint at https://doi.org/10.48550/arXiv.2302.10360 (2023).

8. Xu, Z. et al. Large-scale photonic chiplet Taichi empowers 160-TOPS/W artificial general intelligence. *Science* **384**, 202–209 (2024).

9. Hua, S. et al. An integrated large-scale photonic accelerator with ultralow latency. *Nature* **640**, 361–367 (2025).

10. Ahmed, S. R. et al. Universal photonic artificial intelligence acceleration. *Nature* **640**, 368–374 (2025).

11. Chen, Y. et al. All-optical synthesis chip for large-scale intelligent semantic vision generation. *Science* **390**, 1259–1265 (2025).

12. Kalinin, K. P. et al. Analog optical computer for AI inference and combinatorial optimization. *Nature* **645**, 354–361 (2025).

13. Lin, X. et al. All-optical machine learning using diffractive deep neural networks. *Science* **361**, 1004–1008 (2018).

14. Hu, J. et al. Diffractive optical computing in free space. *Nat. Commun.* **15**, 1525 (2024).

15. Kulce, O., Mengu, D., Rivenson, Y. & Ozcan, A. All-optical information-processing capacity of diffractive surfaces. *Light Sci. Appl.* **10**, 25 (2021).

16. Kulce, O., Mengu, D., Rivenson, Y. & Ozcan, A. All-optical synthesis of an arbitrary linear transformation using diffractive surfaces. *Light Sci. Appl.* **10**, 196 (2021).

17. Liu, C. et al. A programmable diffractive deep neural network based on a digital-coding metasurface array. *Nat. Electron.* **5**, 113–122 (2022).

18. Chen, Y. et al. All-analog photoelectronic chip for high-speed vision tasks. *Nature* **623**, 48–57 (2023).

19. Bernstein, L. et al. Single-shot optical neural network. *Sci. Adv.* **9**, eadg7904 (2023).

20. Fan, Z. et al. Holographic multiplexing metasurface with twisted diffractive neural network. *Nat. Commun.* **15**, 9416 (2024).

21. Luan, C., Davis, R. III, Chen, Z., Englund, D. & Hamerly, R. Single-shot matrix-matrix photonic processor based on spatial-spectral hypermultiplexed parallel diffraction. *Nat. Commun.* **17**, 484 (2026).

22. Shen, Y. et al. Deep learning with coherent nanophotonic circuits. *Nat. Photonics* **11**, 441–446 (2017).

23. Hamerly, R., Bernstein, L., Sludds, A., Soljačić, M. & Englund, D. Large-scale optical neural networks based on photoelectric multiplication. *Phys. Rev. X* **9**, 021032 (2019).

24. Zhu, H. H. et al. Space-efficient optical computing with an integrated chip diffractive neural network. *Nat. Commun.* **13**, 1044 (2022).

25. Feldmann, J. et al. All-optical spiking neurosynaptic networks with self-learning capabilities. *Nature* **569**, 208–214 (2019).

26. Feldmann, J. et al. Parallel convolutional processing using an integrated photonic tensor core. *Nature* **589**, 52–58 (2021).

27. Xu, X. et al. 11 TOPS photonic convolutional accelerator for optical neural networks. *Nature* **589**, 44–51 (2021).

28. Zhang, W., Wang, Y., Lederman, J. C., Shastri, B. J. & Prucnal, P. R. Compact, reconfigurable, and scalable photonic neurons by modulation-and-weighting microring resonators. *eLight* **6**, 6 (2026).

29. LeCun, Y., Bengio, Y. & Hinton, G. Deep learning. *Nature* **521**, 436–444 (2015).

30. Silver, D. et al. Mastering the game of Go with deep neural networks and tree search. *Nature* **529**, 484–489 (2016).

31. Brown, T. B. et al. Language models are few-shot learners. *Adv. Neural Inf. Process. Syst.* **33**, 1877–1901 (2020).

32. Jumper, J. et al. Highly accurate protein structure prediction with AlphaFold. *Nature* **596**, 583–589 (2021).

33. Mennel, L. et al. Ultrafast machine vision with 2D material neural network image sensors. *Nature* **579**, 62–66 (2020).

34. Zhou, T. et al. Large-scale neuromorphic optoelectronic computing with a reconfigurable diffractive processing unit. *Nat. Photonics* **15**, 367–373 (2021).

35. Wang, T. et al. Image sensing with multilayer nonlinear optical neural networks. *Nat. Photonics* **17**, 408–415 (2023).

36. Xue, Z. et al. Fully forward mode training for optical neural networks. *Nature* **632**, 280–286 (2024).

37. Zhang, D. et al. Broadband nonlinear modulation of incoherent light using a transparent optoelectronic neuron array. *Nat. Commun.* **15**, 2433 (2024).

38. Huang, Z. et al. Pre-sensor computing with compact multilayer optical neural network. *Sci. Adv.* **10**, eado8516 (2024).

39. Ning, Y. M. et al. Multilayer nonlinear diffraction neural networks with programmable and fast ReLU activation function. *Nat. Commun.* **16**, 10332 (2025).

40. Xi, H. et al. Ultrasensitive 2D vermiculite inorganic liquid crystals for nonlinear optical activation. *ACS Nano* **20**, 2344–2352 (2026).

41. Song, Y. et al. Integrated electro-optic digital-to-analogue link for efficient computing and arbitrary waveform generation. *Nat. Photonics* **19**, 1107–1115 (2025).

42. Zuo, Y. et al. All-optical neural network with nonlinear activation functions. *Optica* **6**, 1132–1137 (2019).

43. Shen, J. et al. A phase-transition-driven all-optical neuron with sub-nanosecond nonlinear activation. *Adv. Mater.* e22820 (2026).

44. Yan, T. et al. Fourier-space diffractive deep neural network. *Phys. Rev. Lett.* **123**, 023901 (2019).

45. Teğin, U., Yıldırım, M., Oğuz, İ., Moser, C. & Psaltis, D. Scalable optical learning operator. *Nat. Comput. Sci.* **1**, 542–549 (2021).

46. Liu, R. et al. Femto-joule threshold reconfigurable all-optical nonlinear activators for picosecond pulsed optical neural networks. *Light Sci. Appl.* **15**, 128 (2026).

47. Wang, H. et al. Large-scale photonic computing with nonlinear disordered media. *Nat. Comput. Sci.* **4**, 429–439 (2024).

48. Fu, W. et al. Passive all-optical nonlinear neuron activation via PPLN nanophotonic waveguides. *eLight* **6**, 9 (2026).

49. Hazan, A. et al. MXene-nanoflakes-enabled all-optical nonlinear activation function for on-chip photonic deep neural networks. *Adv. Mater.* **35**, 2210216 (2023).

50. Ji, K., Tirabassi, G., Masoller, C., Ge, L. & Yacomotti, A. M. Photonic neuromorphic computing using symmetry-protected zero modes in coupled nanolaser arrays. *Nat. Commun.* **16**, 9203 (2025).

51. Dong, Y. et al. Scalable multilayer diffractive neural network with all-optical nonlinear activation. Preprint at https://doi.org/10.48550/arXiv.2504.13518 (2025).

52. Yan, T. et al. A complete photonic integrated neuron for nonlinear all-optical computing. *Nat. Comput. Sci.* **5**, 1202–1213 (2025).

53. Onodera, T. et al. Arbitrary control over multimode wave propagation for machine learning. *Nat. Phys.* **22**, 164–171 (2026).

54. Xia, F. et al. Nonlinear optical encoding enabled by recurrent linear scattering. *Nat. Photonics* **18**, 1067–1075 (2024).

55. Yildirim, M., Dinc, N. U., Oguz, I., Psaltis, D. & Moser, C. Nonlinear processing with linear optics. *Nat. Photonics* **18**, 1076–1082 (2024).

56. Xu, W. et al. On-chip input-hidden-layer-degenerate optical diffractive nonlinear neural network. *Optica* **13**, 172–180 (2026).

57. Liu, B. et al. Nonlinear optical extreme learner via data reverberation with incoherent light. *Sci. Adv.* **12**, eaeb4237 (2026).

58. Rahman, M. S. S., Li, Y., Yang, X., Chen, S. & Ozcan, A. Massively parallel and universal approximation of nonlinear functions using diffractive processors. *eLight* **5**, 32 (2025).

59. Li, Y., Li, J. & Ozcan, A. Nonlinear encoding in diffractive information processing using linear optical materials. *Light Sci. Appl.* **13**, 173 (2024).

60. Harris, C. R. et al. Array programming with NumPy. *Nature* **585**, 357–362 (2020).

61. Dosovitskiy, A. et al. An image is worth 16×16 words: transformers for image recognition at scale. in *International Conference on Learning Representations* (2021).

62. Jouppi, N. P. et al. In-datacenter performance analysis of a tensor processing unit. in *Proceedings of the 44th Annual International Symposium on Computer Architecture* 1–12 (ACM, 2017).

63. Choquette, J., Gandhi, W., Giroux, O., Stam, N. & Krashinsky, R. NVIDIA A100 Tensor Core GPU: performance and innovation. *IEEE Micro* **41**, 29–35 (2021).

64. Wright, L. G. et al. Deep physical neural networks trained with backpropagation. *Nature* **601**, 549–555 (2022).

65. Momeni, A., Rahmani, B., Malléjac, M., Del Hougne, P. & Fleury, R. Backpropagation-free training of deep physical neural networks. *Science* **382**, 1297–1303 (2023).

66. Momeni, A. et al. Training of physical neural networks. *Nature* **645**, 53–61 (2025).

67. Pai, S. et al. Experimentally realized in situ backpropagation for deep learning in photonic neural networks. *Science* **380**, 398–404 (2023).

68. Bandyopadhyay, S. et al. Single-chip photonic deep neural network with forward-only training. *Nat. Photonics* **18**, 1335–1343 (2024).

69. LeCun, Y., Bottou, L., Bengio, Y. & Haffner, P. Gradient-based learning applied to document recognition. *Proc. IEEE* **86**, 2278–2324 (1998).

70. Xiao, H., Rasul, K. & Vollgraf, R. Fashion-MNIST: a novel image dataset for benchmarking machine learning algorithms. Preprint at https://doi.org/10.48550/arXiv.1708.07747 (2017).

71. Krizhevsky, A. *Learning Multiple Layers of Features from Tiny Images*. Technical Report, University of Toronto (2009).

72. Howard, J. Imagenette: a smaller subset of 10 easily classified classes from ImageNet. https://github.com/fastai/imagenette (2019).

73. Defferrard, M., Benzi, K., Vandergheynst, P. & Bresson, X. FMA: a dataset for music analysis. in *Proceedings of the 18th International Society for Music Information Retrieval Conference* 293–300 (2017).

74. Yu, J. et al. CoCa: contrastive captioners are image-text foundation models. Preprint at

https://doi.org/10.48550/arXiv.2205.01917 (2022).

75. He, K., Zhang, X., Ren, S. & Sun, J. Deep residual learning for image recognition. in *Proceedings of the IEEE Conference on Computer Vision and Pattern Recognition* 770–778 (IEEE, 2016).

76. Wang, Z. et al. Streamlined optical training of large-scale modern deep learning architectures with direct feedback alignment. *Proc. Natl Acad. Sci. USA* **123**, e2532022123 (2026).

77. Wu, T., Li, Y., Ge, L. & Feng, L. Field-programmable photonic nonlinearity. *Nat. Photonics* **19**, 725–732 (2025).

78. Yanagimoto, R. et al. Programmable on-chip nonlinear photonics. *Nature* https://doi.org/10.1038/s41586-025-09620-9 (2025).

79. Ríos, C. et al. Integrated all-photonic non-volatile multi-level memory. *Nat. Photonics* **9**, 725–732 (2015).

80. Chen, Y. et al. Photonic unsupervised learning variational autoencoder for high-throughput and low-latency image transmission. *Sci. Adv.* **9**, eadf8437 (2023).

81. Chen, S., Li, Y., Wang, Y., Chen, H. & Ozcan, A. Optical generative models. *Nature* **644**, 903–911 (2025).

# Methods

## Dataset preparation

**MNIST dataset.** The MNIST dataset contains handwritten digits from ten categories, comprising 60,000 training samples and 10,000 test samples. Each grayscale image has a resolution of 28 × 28 pixels. Images were normalized to the range [0, 1] and encoded as amplitude patterns on the SLM, with phase values set to zero. This encoding allows the direct projection of digit patterns onto the SLM for all-optical inference without electronic preprocessing.

**Fashion-MNIST dataset.** The Fashion-MNIST dataset comprises ten categories of fashion products, consisting of 60,000 training samples and 10,000 test samples. Each grayscale image has a resolution of 28 × 28 pixels and is normalized identically to the MNIST preprocessing pipeline. Processed images are encoded as amplitude patterns on the SLM with uniform phase distributions.

**CIFAR-10 dataset.** The CIFAR-10 dataset contains 50,000 training images and 10,000 test images distributed across ten natural object categories. Each 32 × 32 pixels RGB image is converted to 3 grayscale channels to maintain single-channel optical encoding. Amplitude-encoded samples are projected into the optical network with a constant phase baseline. This dataset is employed to evaluate the network's generalization capability on natural scenes with greater visual complexity.

**Imagenette dataset.** The Imagenette dataset is a subset of ImageNet containing ten easily classified categories. Each RGB image is converted to 3 grayscale channels, resized to 32 × 32 pixels, and normalized before encoding. Optical fields corresponding to these images are directly modulated on the SLM and propagated through the ONNPU system for inference.

**Geometric Kakeya dataset.** A synthetic geometric dataset is designed to evaluate abstract geometric reasoning capabilities. The dataset comprises binary images representing four categories: convex Kakeya, concave Kakeya, convex non-Kakeya, and concave non-Kakeya geometries. Each binary mask is resized to 32 × 32 pixels, smoothed with a Gaussian kernel, and encoded as amplitude patterns on the SLM with zero phase. These patterns test the network's ability to learn global geometric properties relevant to rotational containment, rather than local texture recognition.

**Music genre dataset.** The FMA dataset is used for music genre classification, comprising audio samples from 8 musical genres: Rock, Experimental, International, Instrumental, Hip-hop, Folk, Pop and Electronic music. Each audio clip is converted into a Mel-spectrogram representation, which is then normalized and resized to 32 × 32 pixels to match the optical input plane. These spectrograms are amplitude-encoded as two-dimensional optical intensity patterns, with phase values initialized to zero. Each spectrogram represents a distinct optical input corresponding to the time-frequency distribution of music features.

**Regression datasets.** Three regression tasks are designed to evaluate continuous function fitting capabilities. The first task involves fitting a discontinuous step function, inspired by industrial inspection applications detecting sudden state transitions. The second task targets a localized Gaussian-like peak function, relevant to target acquisition in radar systems. The third task involves fitting a rapidly oscillating periodic waveform with sharp spikes, applicable to medical examination monitoring periodic physiological signals. Each target function is discretized into 5,000 points, normalized, and encoded as amplitude patterns for optical processing.

**ImageNet dataset.** The ImageNet dataset is organized according to the WordNet hierarchy, containing over 1.2 million training images and 50,000 validation images across 1,000 object categories. For the large-scale classification experiments, each image is converted to 3 grayscale channels, resized to 256 × 256 pixels to match the Vision Transformer input requirements, and normalized. Images are encoded as amplitude patterns on the SLM with zero phase for optical processing through the hybrid ONNPU ViT architecture.

## Training methods

**Training of the all-optical configuration.** For tasks on MNIST, Fashion-MNIST and Geometric Kakeya datasets, the ONNPU was trained in an all-optical configuration where the entire inference pipeline operates in the optical domain without electronic post-processing. The diffractive phase profiles were optimized using standard backpropagation with an optical forward model implemented in PyTorch. The optical forward model simulated complex-field evolution through cascaded diffractive layers and self-rectification nonlinear activation. The training objective minimized the cross-entropy loss between the predicted optical intensity distribution on the output detector and the target class regions. The diffractive layer parameters were updated using the Adam optimizer with a learning rate of 0.001 and a batch size of 128. Training converged typically within 30-50 epochs. Experimental errors, including geometric misalignment errors in *X*/*Y*/*Z* translations and rotations, as well as phase fabrication errors in the diffractive optical elements, were all prospectively included during training to ensure robustness (Supplementary Information, Note S6 and Fig. S1). The optimized phase distributions were then uploaded to the DOE phase masks for experimental validation. Classification decisions were determined by identifying the detector region with maximum optical intensity.

**Training with electronic readout.** For RGB datasets (CIFAR-10, Imagenette) and datasets requiring additional encoding or decoding (Music genre, regression), a hybrid training approach was employed where 8 ONNPU modules perform cascaded linear computation and nonlinear processing operations. Each ONNPU module in the forward pass processes the input through diffractive layers and self-rectification activation, with the output from one module serving as input to the next. This multi-module computational pipeline is followed by a lightweight linear electronic readout layer (a single fully connected layer with fewer than $10^3$ trainable parameters) for final classification. The optical modules were trained using backpropagation through the optical forward model, with gradients computed and propagated through each ONNPU module in the backward pass, while the electronic readout layer was optimized simultaneously. The loss function combined the optical feature representation from the multi-module pipeline with the electronic classifier output, enabling end-to-end optimization of the hybrid system. Training was performed with the Adam optimizer using a learning rate of 0.0005 and batch size of 64, converging within 50-75 epochs.

**Training ONNPU Vision Transformer.** For large-scale ImageNet classification with 1,000 categories, multiple ONNPU modules were integrated into a ViT architecture. The ONNPU modules replaced the computationally intensive linear transformations and nonlinear activations in the embedding, transformer encoder, and classifier stages, while electronic computing was retained for positional encoding, attention matrix operations ($QK^\top$), and softmax normalization. The training process involved pre-training the entire hybrid ONNPU ViT architecture on ImageNet data, followed by fine-tuning the ONNPU modules. The optical forward model accurately simulated the self-rectification nonlinear activation and diffractive propagation, enabling gradient computation through the optical layers. The training objective minimized cross-entropy loss between predicted and ground-truth class labels. Optimization was performed using the AdamW optimizer with a learning rate of 0.0001, weight decay of 0.01, and batch size of 32. The model was trained for 75 epochs, achieving 82.4% top-1 accuracy and 96.5% top-5 accuracy on the ImageNet validation set.

## Experiments

**Activation function characterization.** The self-rectification nonlinear activation was characterized using an optical test bench (Supplementary Information, Fig. S12). Femtosecond pulses from an erbium-doped fiber laser (FSL(Er)780/1550) with a center wavelength of 1560 ± 10 nm, pulse width ≤150 fs, repetition rate of 100 MHz, maximum power of 1 W, and pulse energy of 10.0 nJ were collimated and expanded. The beam was relayed through two lenses to control the beam diameter and focus inside the nonlinear module. The self-rectification module consisted of a β-$BaB_2O_4$ (BBO) crystal with dimensions of 20 mm × 20 mm × 0.9 mm, cut angle $\theta$ = 19.8°, and a low-pass spectral filter with a passband of 400–900 nm. After passing through the self-rectification module, the transmitted light was split by a non-polarizing beam splitter (NPBS, JCOPTIX BS2555-T4M, operating wavelength 1,100–1,600 nm) into two arms and monitored by two calibrated power meters. One arm recorded the incident power on the activation module, while the other arm recorded the transmitted (activated) power at the second-harmonic frequency. By sweeping the input power over the operating range and adjusting the crystal orientation using a precision rotary stage, we obtained the input-output transfer curve of the self-rectification process. The measured activation characteristics closely matched theoretical predictions based on second-harmonic generation, confirming the threshold-like transfer function implemented entirely in the optical domain.

**Free-space experimental system.** All ONNPU experiments were performed in a free-space optical setup built on a vibration-isolated optical table (Fig. 2b–c). The effective aperture of the system was 2 cm, matching the illuminated area on the phase masks. Input data were encoded on a reflective SLM (LBTEK 067DN1) with an operating wavelength range of 750–1630 nm, phase modulation range of 0–2$\pi$, pixel count of 904 × 800, and pixel size of 16.2 µm × 10.8 µm. The diffractive linear layers were implemented by a cascade of transmissive DOEs with an active area of 12.8 mm × 12.8 mm. The optical beam was relayed from the SLM to the DOE stack by a telescope system comprising two achromatic lenses (working wavelength 1,050–1,700 nm): the first lens with a diameter of 12.7 mm and focal length of 25.0 mm, and the second lens with a diameter of 25.4 mm and focal length of 100.1 mm. A half-wave plate (working wavelength 1550 nm) was used to control the polarization state. Dielectric mirrors with silver coatings providing high reflectivity over a broad wavelength range (450 nm–20.0 µm) were used for beam steering. Between the second and third diffractive layers, the self-rectification module was inserted to provide all-optical nonlinear activation. Output patterns were imaged using a CMOS camera (AIC-501GM-GE, GigE monochrome camera with CMOS sensor, 2448 × 2048 pixels, pixel size 3.45 µm × 3.45 µm).

## Data availability

All data required to evaluate the conclusions of this study are presented in the Article or its Supplementary Information. All relevant code is available from the corresponding author upon reasonable request. The data repository is further available via Dryad at https://doi.org/10.5061/dryad.fttdz097k. Please note that this DOI is currently reserved and will be formally activated upon the manuscript's publication. For peer-review purposes, the dataset can be accessed immediately via the following temporary link: http://datadryad.org/share/6bVcKigRNnJASc0Igxa-xekiOCHl-JU7wIGxBhKocwQ.

## Acknowledgements

This work was supported by National Key Research and Development Program of China (2023YFF0613600), National Natural Science Foundation of China (62475192, 62205246, 61621001, 62192770, 62192772, 12274296, 62020106009 and 62111530053), Shanghai Pilot Program for Basic Research, Science and Technology Commission of Shanghai Municipality (17JC1400800, 20JC1414600 and 21JC1406100), the "Shu Guang" project supported by Shanghai Municipal Education Commission and Shanghai Education (17SG22), and Fundamental Research Funds for the Central Universities.

## Author contributions

Y.S. and R.M. initiated the project. R.M. conceived the idea, designed the computing systems and trained neural networks. R.M. and Y.Z. constructed the experimental system and conducted the task validations. S.D., Q.F., H.A., W.H. and X.D. fabricated the samples. H.L. provided technical support in experiment. All authors analyzed the results. R.M. and Y.S. wrote the manuscript. R.F. and A.M. revised the paper. Y.S., Z.W., R.F. and X.C. supervised the project.

## Competing interests

The authors declare no competing interests.

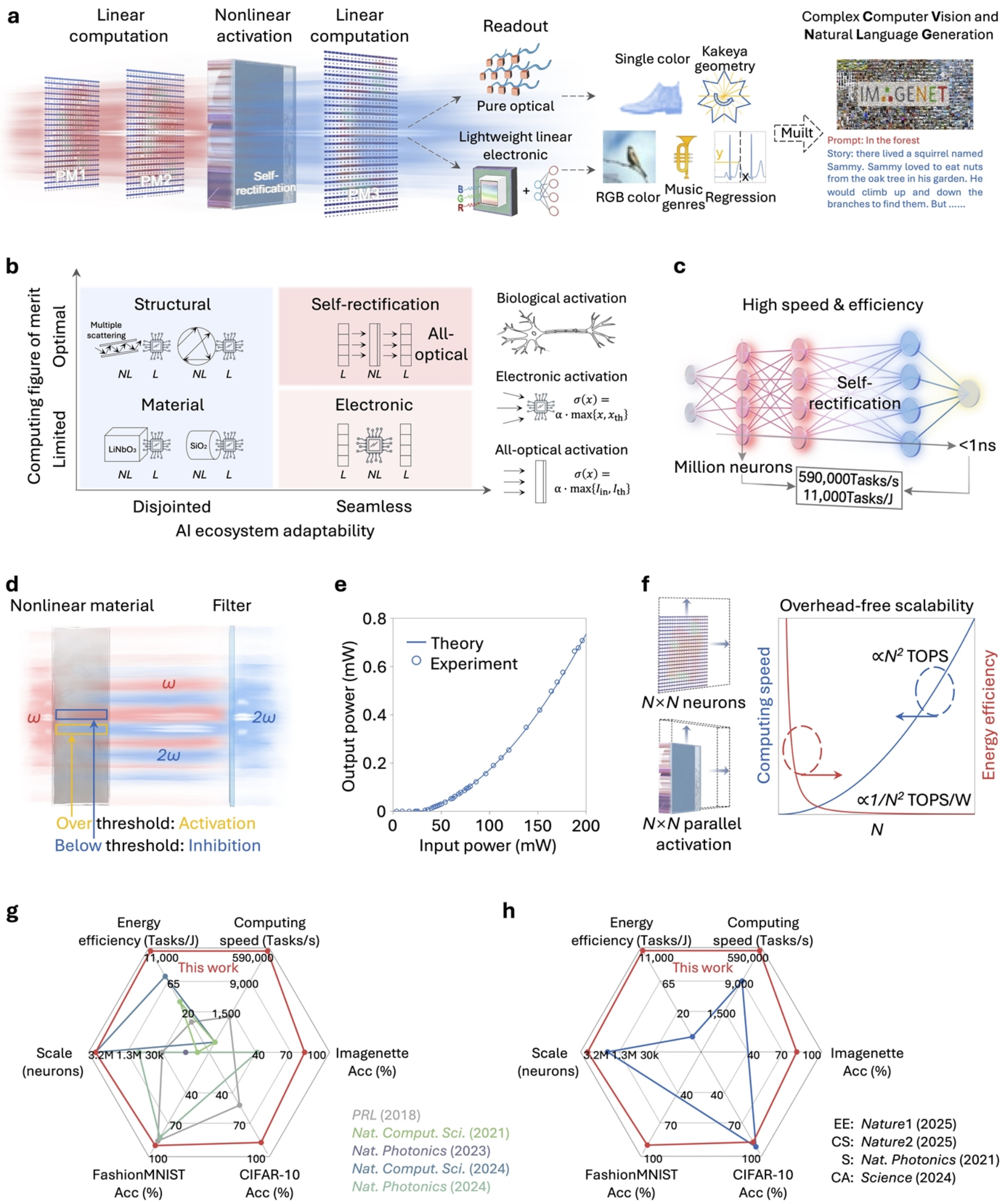

**Fig. 1: Architecture and performance of ONNPU. (a)** ONNPU implements the canonical "linear transform–nonlinear activation–linear transform" neural-network structure using cascaded diffractive phase modulators interleaved with an all-optical self-rectification activation module, with readout that can be purely optical or augmented by a lightweight linear electronic layer. An all-optical configuration is used for MNIST, FashionMNIST and Kakeya geometry classification. For CIFAR-10, Imagenette, music genres classification and regression, we add a lightweight linear electronic readout. **(b)** Positioning of nonlinear optical computing approaches by computing FoM (figure of merit, factoring in processing speed, energy

efficiency, and hardware scalability) and architectural adaptability to AI ecosystem; self-rectification provides optimal nonlinearity while preserving a seamless architecture. *NL*, nonlinear computing; *L*, linear computing. **(c)** Network representation of ONNPU showing all-to-all optical connectivity between layers with an embedded self-rectification nonlinearity. **(d)** Operating principle of the self-rectification mechanism. Incident light at frequency $\omega$ passes through a nonlinear material; regions with intensity above the activation threshold generate second-harmonic light at $2\omega$ (activation), while sub-threshold regions remain at $\omega$ (inhibition). A bandpass filter blocks the fundamental frequency, transmitting only the activated $2\omega$ signal. **(e)** Measured activation characteristic showing output power versus input power, with experimental data closely matching the simulation. **(f)** Scaling properties: computing speed scales quadratically with neuron count ( $\propto N^2$ TOPS) while latency remains constant. **(g)** Radar chart comparing ONNPU with other electro-optic hybrid and optical nonlinear optical neural networks across six metrics: computing power (tasks/s), energy consumption (tasks/J), scale (neurons), and classification accuracy on Fashion-MNIST, CIFAR-10 and Imagenette. **(h)** Radar chart comparing ONNPU with state-of-the-art electronic-nonlinear optical neural networks on the same metrics.

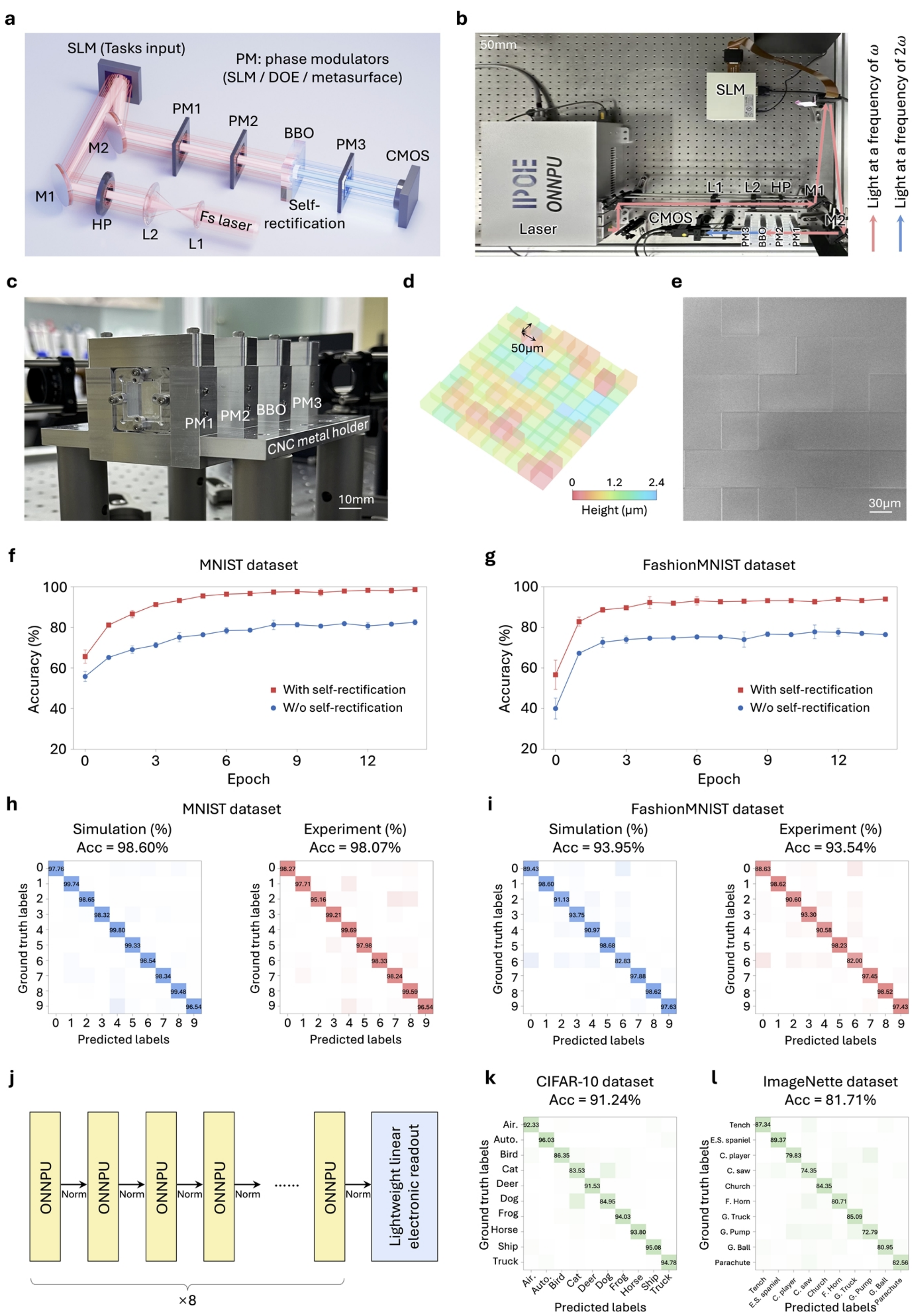

**Fig. 2: Experimental configuration of ONNPU and benchmarking results on image classification tasks. (a)–(c)** Experimental pipeline with an SLM as input, phase modulators

(SLM/DOE/metasurface), a self-rectification module, and a CMOS camera. L1, L2, lens; HP, half wave plates; M1, M2, mirrors; PM1, PM2, PM3, DOE (diffractive optical element) phase masks. **(d)** 3D surface profile of DOE unit cells measured by a stylus profiler. The scanning area is 500 µm × 500 µm. **(e)** SEM image of the DOE. Scale bar, 30 µm. **(f–g)** Training curves comparing networks with self-rectification (yellow) versus without self-rectification (blue) on MNIST **(f)** and Fashion-MNIST **(g)**. Self-rectification enables rapid convergence to high accuracy, while linear networks plateau at lower performance. **(h)** Simulation and experiment classification results on MNIST (98.07% accuracy) **(i)** Simulation and experiment classification results on FashionMNIST (93.54% accuracy). **(j)** Hybrid multi-pass configuration followed by a lightweight linear electronic readout for higher-complexity tasks. The demonstrations are conducted based on an error-aware optical forward model. **(k)–(l)** Classification results on more complex datasets. **(k)** Confusion matrix for CIFAR-10 (91.24% accuracy) with 10 object categories. **(l)** Confusion matrix for Imagenette, an ImageNet subset (81.71% accuracy) comprising 10 categories.

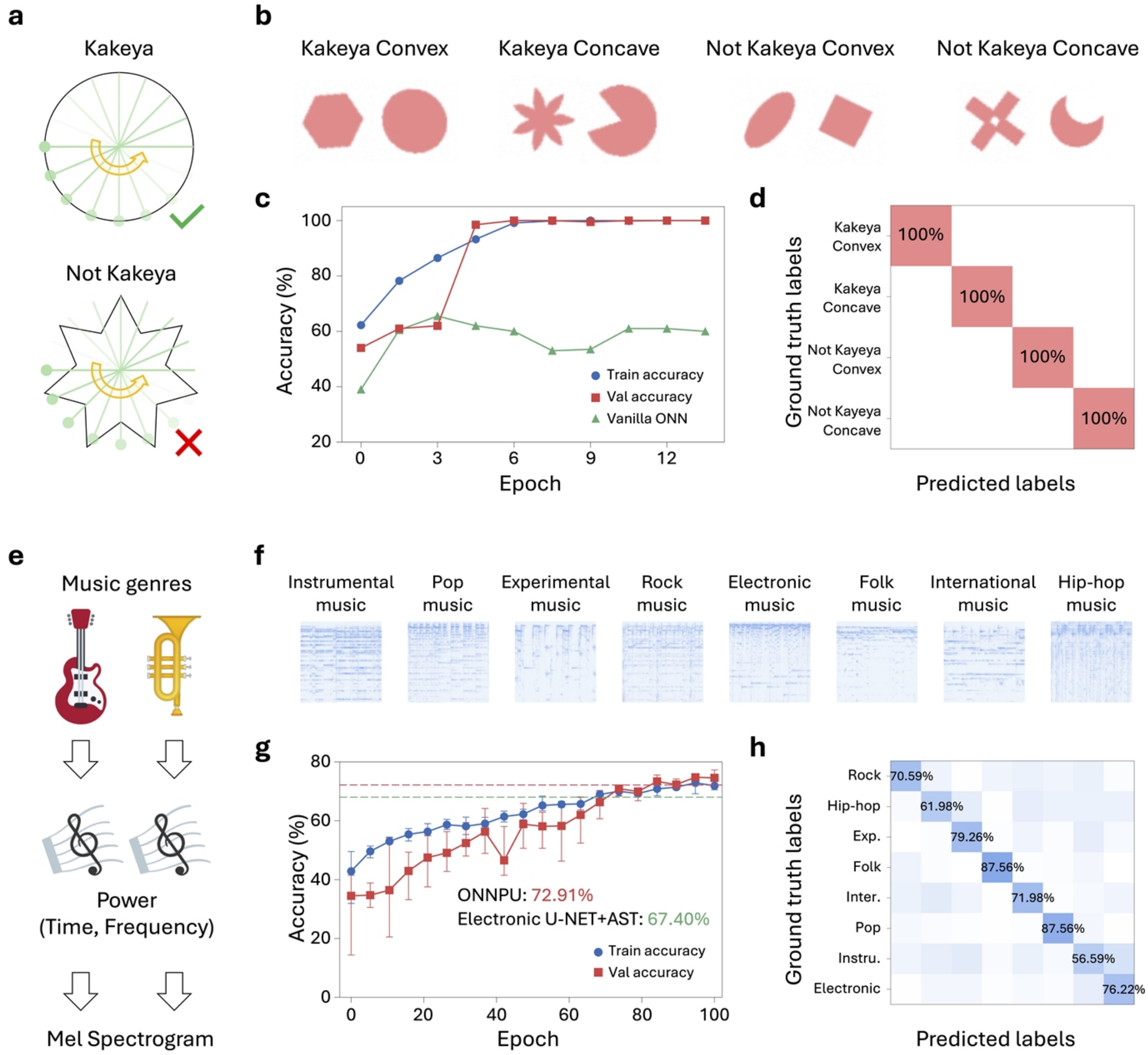

**Fig. 3: Performance of error-aware optical forward model on diverse input decision tasks.** **(a)** Illustration of the Kakeya membership problem: a shape belongs to the Kakeya set if a unit line segment can rotate 360° within it; otherwise, it does not. **(b)** Four input shape classes: Kakeya & Convex, Kakeya & Concave, Not Kakeya & Convex, Not Kakeya & Concave. **(c)** Training curves for geometry classification showing accuracy and loss over 15 epochs, achieving near-100% accuracy. The demonstration is conducted based on an error-aware optical forward model. **(d)** Confusion matrix for geometry recognition. **(e)** Encoding pipeline for music genre classification: audio waveforms are converted to Mel-spectrograms and encoded as 2D optical intensity patterns. **(f)** Representative Mel-spectrogram inputs for 8 musical genres. **(g)** Training curves for music genre classification showing accuracy as a function of training epoch (up to 100). The demonstration is conducted based on an error-aware optical forward model. **(h)** Confusion matrix for music genres recognition.

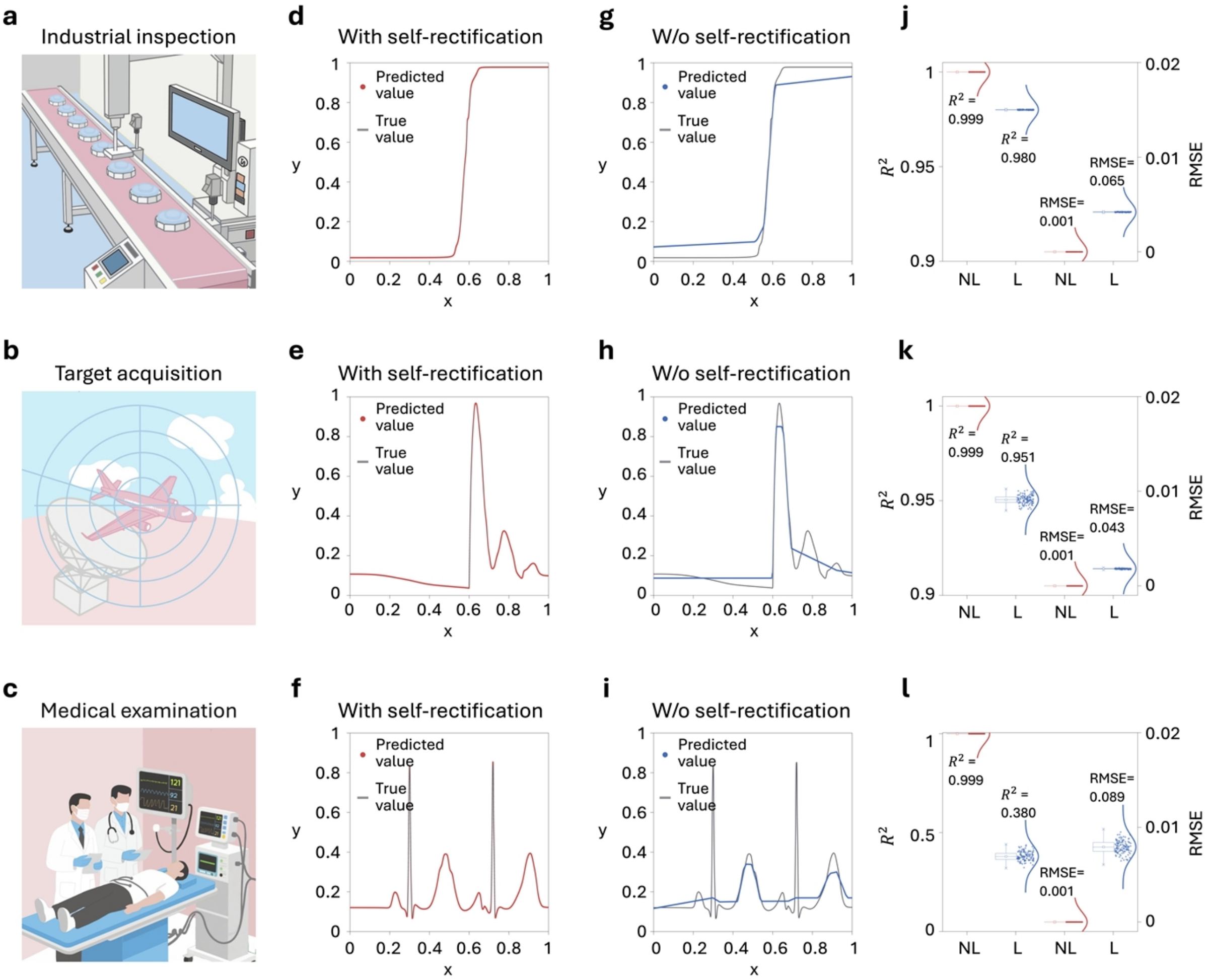

**Fig. 4: Regression capabilities of ONNPU. (a)–(c)** Three real-world application scenarios motivating the regression tasks: **(a)** industrial inspection on a conveyor belt requiring detection of state transitions, **(b)** target acquisition in radar/tracking systems requiring localization of events, and **(c)** medical examination requiring monitoring of periodic physiological signals. **(d)–(f)** Representative fitting results with self-rectification, showing predicted curves (red) closely tracking the ground truth (grey) for three nonlinear target functions. The demonstrations are conducted based on an error-aware optical forward model. **(g)–(i)** Corresponding fits without self-rectification (linear optical network, blue), showing degraded fidelity and failure to capture discontinuities, sharp peaks and high-frequency structure. **(j)–(l)** Summary statistics of regression performance ($R^2$ and RMSE) comparing nonlinear (NL; nonlinear, with self-rectification) and linear (L; linear, without self-rectification) models for the three tasks.

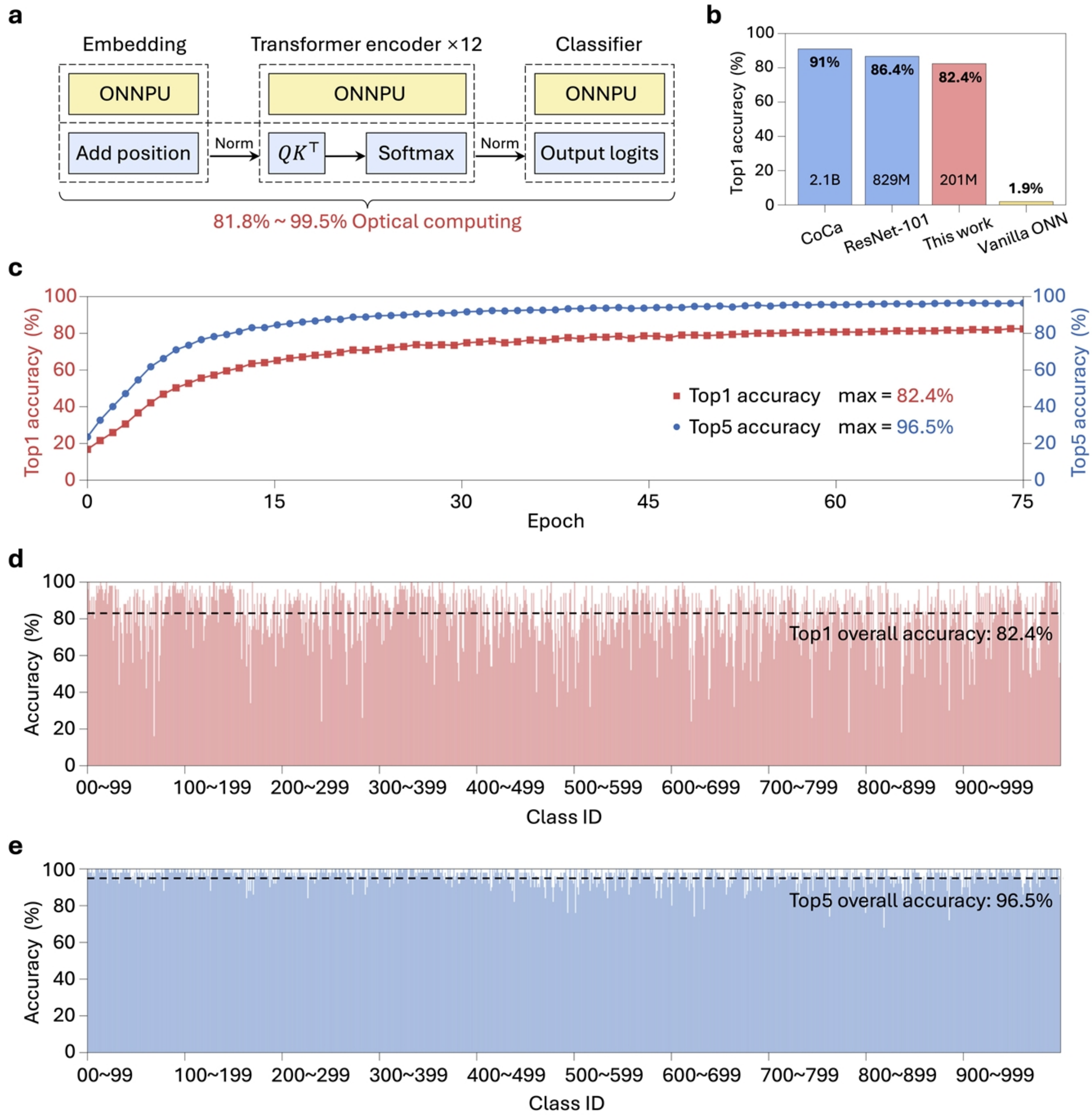

**Fig. 5: Performance of error-aware optical forward model on full ImageNet classification. (a)** ONNPU Vision Transformer architecture showing integration of ONNPU modules (orange boxes) for linear transformations and nonlinear activations in the embedding, transformer encoder, and classifier stages, while retaining electronic computing (light green boxes) for position encoding, attention matrix operations ($QK^{\top}$), and softmax normalization. 81.8% to 99.5% of the system's computation is performed in the optical domain (See Supplementary Information, Note S8 and Fig. S6 for details). **(b)** Top-1 accuracy comparison of ONNPU ViT (82.4% with 201M parameters, red) with other models, including CoCa (91.0% with 2.1B parameters), ResNet-101 (86.4% with 829M parameters), and vanilla optical neural network (1.9%). The demonstration is conducted based on an error-aware optical forward model. **(c)** Training curves showing top-1 accuracy (red, left axis) and top-5 accuracy (blue, right axis) over 75 epochs on the full ImageNet dataset with 1,000 object categories. **(d)** Per-class top-1 accuracy distribution across all 1,000 ImageNet categories, showing variation in recognition

difficulty among different object classes. **(e)** Per-class top-5 accuracy distribution, most classes exceeding 95% accuracy.

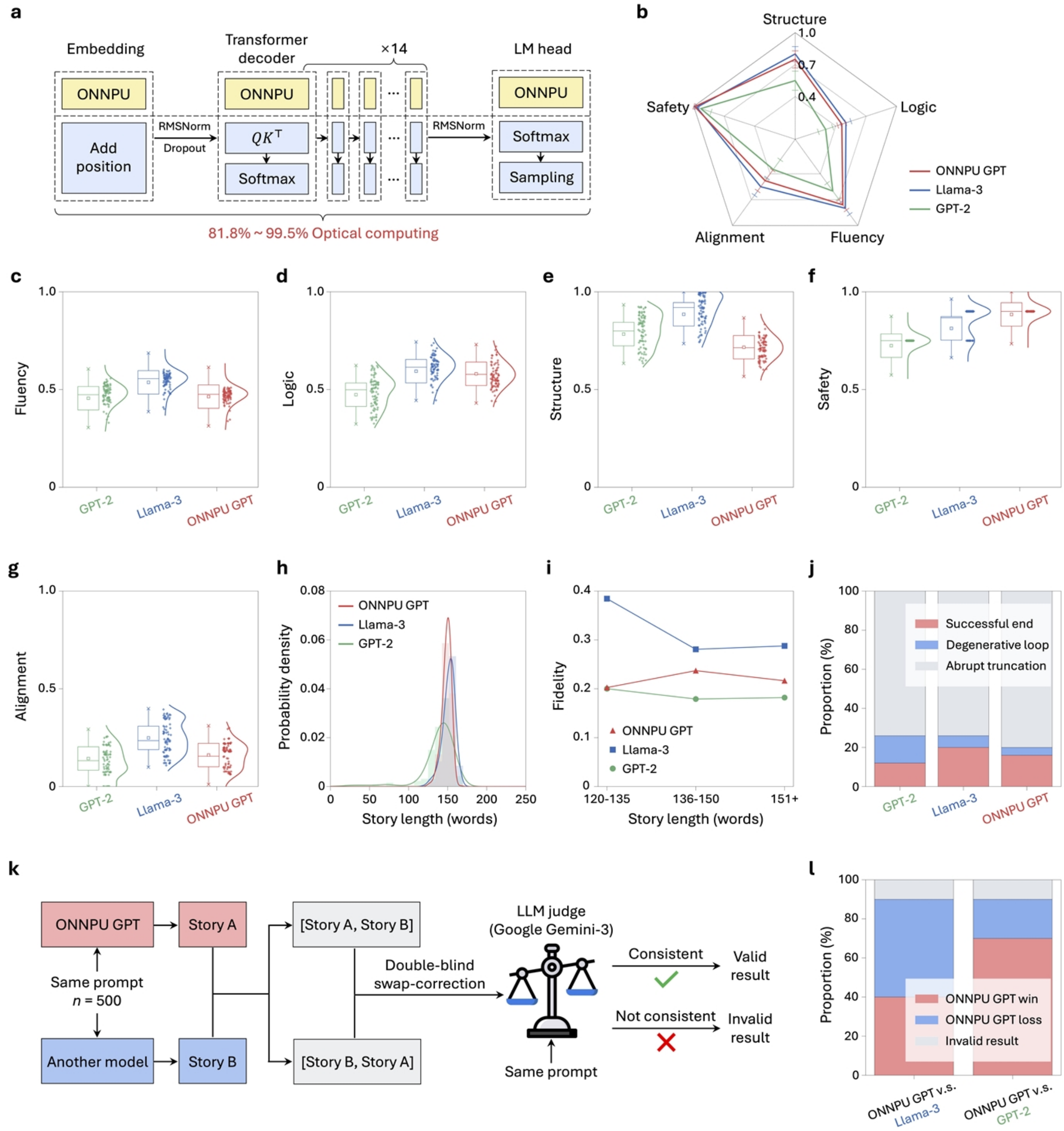

**Fig. 6: Language generation with a decoder-only, causal Transformer implemented with ONNPU modules (ONNPU GPT). (a)** ONNPU GPT architecture: ONNPU modules implement the dominant linear projections and nonlinear activations throughout the token embedding, 14 Transformer decoder blocks and the language-model head, while electronic computation is retained for RMSNorm, dropout, direct attention matrix multiplication ($QK^\top$), softmax and sampling, resulting in 81.8% to 99.5% optical computing. (See Supplementary Information, Note S8 and Fig. S7 for details) **(b)** Radar-plot summary of language quality scores (Supplementary Information, Note S9) across structure, logic, fluency, alignment and safety for ONNPU GPT (117M parameters) compared with electronic baselines (GPT-2, 124M; Llama-3, 8B). The demonstration is conducted based on an error-aware optical forward model. **(c)–(g)** Distributions of language quality scores for individual criteria (fluency, logic, structure, safety

and alignment) across the evaluation set. **(h)** Story-length distributions of generated continuations. **(i)** Fidelity as a function of story length. **(j)** End-type of generated stories, categorized as successful completion, degenerative loop or abrupt truncation. **(k)** Double-blind swap-correction (DBSC) evaluation protocol: for each story (total number $n$ = 500), ONNPU GPT and a baseline model generate a continuation (Story A and Story B); paired outputs are swap-corrected and scored by an independent LLM judge (Google Gemini-3) to determine the superior output. **(l)** Head-to-head preference statistics of ONNPU GPT versus electronic baselines, reporting win/loss/invalid proportions under the same judging protocol.

| Reference | Nonlinear method | Tasks performed | Computing Speed (Tasks/s) | Energy Efficiency (Tasks/J) | Computing Speed (TOPS) | Energy Efficiency (TOPS/W) | Neuron number |
|---|---|---|---|---|---|---|---|
| 2D-material sensor *Nature*, 2020 (33) | Electronic | 10-category MNIST (92.7%) | NA | NA | 0.001 | NA | 27 |
| DPU *Nature Photonics*, 2021 (34) | Electronic | 10-category MNIST (96.6%), 10-category Fashion-MNIST (84.6%), High-speed image and video recognition | ~145 | ~0.91 | 240.1 | 1.578 | $2.2\times10^{6}$ |
| PACE *Nature*, 2025 (9) | Electronic | Computationally hard Ising problems | NA | NA | 8.19 | 2.38 | $4.1\times10^{3}$ |
| Pre-sensor encoder *Nature Photonics*, 2023 (35) | Electro-optic hybrid | 10-category QuickDraw, Cell-organelle classification, 3D real-scene objects recognition | 10 | 1.54 | NA | NA | 1,640 |
| Theoretical maximum | | NA | $1.6\times10^{4}$ | $1.78\times10^{4}$ | NA | NA | |
| AONN *Optica*, 2019 (42) | Material (laser-cooled atoms) | Ising model phase classification | NA | NA | $\sim8\times10^{-10}$ | $\sim8\times10^{-12}$ | 22 |
| F-D2NN *PRL*, 2019 (44) | Material (ferroelectric thin films) | 10-category MNIST (98.1%), Cell segmentation, Video salient object detection | ~60 | ~0.12 | $\sim1\times10^{-4}$ | ~500 | $3.2\times10^{6}$ |
| SOLO *Nature Computational Science*, 2021 (45) | Material (multimode $SiO_2$ fibers) | 2-category COVID-19 X-ray, Regression | 2 | NA | $1\times10^{-7}$ | 105 | 240 |
| Disordered *$LiNbO_3$* media *Nature Computational Science*, 2024 (47) | Material (disordered $LiNbO_3$) | 10-category MNIST (96.48%), 10-category Fashion-MNIST (87.58%), 10-category CIFAR (52.2%), SLD & ASL classification, Regression, Graph classification | 40 | 0.54 | 18.8 | 37.23 | $3.12\times10^{4}$ |
| nPOLO *Nature Photonics*, 2024 (55) | Structural | 10-category MNIST (88.0%), 10-category Fashion-MNIST (84%), 10-category ImageNet subset (41.10%) | ~60 | ~0.44 | NA | NA | $3.6\times10^{5}$ |
| Multiple-scattering cavity *Nature Photonics*, 2024 (54) | Structural | 10-category MNIST (83.5%), Image reconstruction, Keypoint detection, Real-time video pedestrian detection | ~60 | ~0.19 | NA | NA | $4.1\times10^{6}$ |
| Google TPU (62) | NA | General purpose computing | $\sim1.12\times10^{4}$ (ResNet-50) | ~280 (ResNet-50) | 92 | 2.3 | NA |
| NVIDIA GPU A100 (63) | NA | General purpose computing | $\sim1.9\times10^{4}$ (ResNet-50) | ~63.4 (ResNet-50) | 156 | 0.52 | NA |
| **ONNPU this work** | Self-rectification | 10-category MNIST (98.07%),<br>10-category Fashion-MNIST (93.54%),<br>geometry recognition | $5.90\times10^{5}$ | $1.10\times10^{4}$ | 0.727 | $1.35\times10^{-2}$ | $3.23\times10^{6}$ |
| | | 10-category CIFAR (91.24%),<br>10-category Imagenette (81.71%),<br>music genres recognition,<br>Regression | $5.65\times10^{4}$ | $1.05\times10^{3}$ | 0.949 | $1.76\times10^{-2}$ | $2.58\times10^{7}$ |
| | | 1000-category ImageNet (82.4%) | 504 | 9.36 | 60.7 | 1.13 | $2.01\times10^{8}$ |
| **With advanced SLM** | | NA | $2.11\times10^{7}$ | $3.91\times10^{5}$ | $7.13\times10^{7}$ | $1.32\times10^{6}$ | NA |
| **Theoretical maximum** | | NA | NA | NA | $6.36\times10^{9}$ | $1.18\times10^{8}$ | NA |

**Table 1: Performance of ONNPU compared with state-of-the-art electronic and photonic devices.** Our reported TOPS and TOPS/W values are lower than those in some existing literature due to a difference in counting methodology. We adopt a task-centric approach (considering the specific input size for each task and reserved guarded spacing to ensure robust signal separation), while some conventions are based on the full hardware bandwidth (counting operations across the entire SLM aperture). To ensure a more robust and equitable end-to-end comparison, we additionally report tasks/s and tasks/J as direct metrics, which strictly reflect functional throughput and system efficiency. The evaluation methodology for this work is detailed in Supplementary Information, Note S1 and Table S1. Values marked with "~" are theoretically calculated from reported system parameters. Structural nonlinearity, nonlinear input–output behavior achieved in linear optical systems by design (e.g. repeated encoding, multiple scattering or repeated passes through linear optics), so that the overall mapping becomes a nonlinear function of the input, typically a polynomial.